\documentclass[preprint,showpacs,amsmath,amssymb]{revtex4-1}
\usepackage{graphics}
\usepackage{graphicx}
\usepackage{dcolumn}
\usepackage{epsfig}
\usepackage{bm}
\newcommand{\be}{\begin{eqnarray}}
\newcommand{\en}{\end{eqnarray}}
\newcommand{\bes}{\begin{subequations}}
\newcommand{\ens}{\end{subequations}}
\newcommand{\ben}{\begin{eqnarray*}}
\newcommand{\enn}{\end{eqnarray*}}
\newcommand{\beq}{\begin{equation}}
\newcommand{\enq}{\end{equation}}

\newcommand{\f}{\frac}

\newcommand{\mc}[1]{\mathcal{#1}}

\newcommand{\bi}{\begin{itemize}}
\newcommand{\ei}{\end{itemize}}

\renewcommand{\t}{\Theta}
\renewcommand{\theta}{\Theta}

\begin{document}
\title{Renormalization group and isochronous oscillations}
\author{Amartya Sarkar}
\email{amarta345@bose.res.in}
\author{J.K. Bhattacharjee}
\email{jkb@bose.res.in}
\affiliation{Department of Theoretical Sciences, S. N. Bose National Centre for Basic Sciences, Salt lake, Kolkata 700098, India}

\date{\today}
\begin{abstract}
We show how the condition of isochronicity can be studied for two dimensional systems in the renormalization group (RG) context. We find a necessary condition for the isochronicity of the Cherkas and another class of cubic systems. Our conditions are satisfied by all the cases studied recently by Bardet et al \cite{bard} and Ghose Choudhury and Guha
\end{abstract}
\maketitle
\section{Introduction}
Isochronous systems have been quite fascinating ever since Galileo's discovery of first such system - the simple harmonic oscillator. Eminent names such as Newton, Huygens, Jacobi, Poincare etc have delved in the investigation of isochronous systems. In the last century the focus has mainly been on 2-dimensional cases \cite{chasa} - as it is related to the Hilbert's 16th problem. The other paradigm \cite{anbr,amp,calo} of an isochronous system is the rather unexpected case of the potential $V(x)=\frac{1}{2}\left(x^2+\frac{1}{x^2} \right)$. In a two dimensional dynamical system, isochronicity implies that the time period is independent of the amplitude of motion (i.e. independent of the initial condition). An exhaustive study of various isochronous centers of vector fields in the plane can be found in \cite{chasa}. In recent years the seminal work of Calogero and collaborators \cite{calon,calo1,calo2,calo3} has opened up new and interesting directions in the study of isochronous systems. Calogero et al introduced a simple trick by which one can construct isochronous oscillators by modifying equations for ordinary oscillators. And this led to the conclusion that - ``isochronous systems are not rare" \cite{calo4}.This area of research focuses mainly in the construction of isochronous systems and has been widely exploited of late by Ghose Choudhury and Guha \cite{guch}. In this work however we explore a different angle. Given a two dimensional dynamical system we propose to find the conditions for isochronicity by using renormalization group technique.
\\
\section{The Method}
The renormalization group approach has been an effective tool for analyzing periodic orbits in dynamical systems. First introduced by Chen et al for the purpose of extracting asymptotic behavior of the solutions to various differential equations, it has had numerous applications over the following years. Recently it has been demonstrated to be an useful method for differentiating between center like oscillations and limit cycles for two dimensional dynamical systems \cite{asjkbsc,asjkbepl}. In fact, that differentiation has been shown to lead to a classification of orbits in Lienard systems of different orders \cite{dhjkblin} i.e. the dynamical system, $\ddot{x}+\dot{x}^nf(x)+x+g(x)=0$, where $n=1,2,\cdots$. In these Lienard systems, $f(x)$ and $g(x)$ are nonlinear functions of $x$, with $g(x)$ not having any linear term. In this communication we show that the renormalization group can also be effective for identifying isochronous oscillations.
\\
The methodology of implementing the renormalization group is as follows:
\begin{enumerate}
\item Identify a fixed point that is either directly a linear center or can be made a linear center by suitably choosing a parameter in the problem (this parameter may in certain situations have to be introduced).
\item Set up a perturbation theory around the closed orbit of the center. The orbit will be characterized by two constants --- the amplitude $A$ and the phase $\theta$, fixed by the two initial conditions associated with a second order dynamical system.
\item The perturbation theory will most likely diverge (due to the presence of secular terms) as the separation $t-t_0$ ($t$ being the current time and $t_0$ the initial time) becomes large. Two renormalization constants have to be introduced to absorb these divergences from the past ($t_0\rightarrow-\infty$). The renormalization constants appear in terms of an arbitrary time $\tau$, which serves to peg the new initial conditions. The amplitude and phase are now dependent on $\tau$.
\item The value of $x$ at time $t$ (the current time) cannot depend on where one sets the initial conditions and hence $\left(\frac{\partial x}{\partial\tau}\right)_t=0$. This leads to the flow equations.
\end{enumerate}
\begin{eqnarray}
\frac{dA}{d\tau} &=& f(A) \label{4.001}\\
\frac{d\theta}{d\tau} &=& g(A)\label{4.002}
\end{eqnarray}
If the oscillation is of center variety, then the initial condition sets the amplitude of motion and hence ${dA}/{d\tau}=0$. For an isochronous orbit of the center variety 
\beq
f(A)=0;\quad\quad g(A)=0\quad identically. \label{4.003}
\enq
We now present an algorithm for calculating $f(A)$ and $g(A)$. This is best done by considering the anharmonic oscillator
\beq
\ddot{x}+\omega^2 x = -\lambda x^3 \label{4.004}
\enq
The expansion $x=x_0+\lambda x_1+\lambda^2 x_2+\cdots$ leads to the system of equations (different orders of $\lambda^n$)
\be
n=0: \quad\quad \ddot{x}_0+\omega^2 x_0 &=& 0 \label{4.005}\\
n=1: \quad\quad \ddot{x}_1+\omega^2 x_1 &=& -x_0^3 \label{4.006}\\
n=2: \quad\quad \ddot{x}_2+\omega^2 x_2 &=& -3x_0^2x_1 \label{4.007}
\en
The initial condition is taken as $x(t_0)=A$ and $\dot{x}(t_0)=0$. The solution at $n=0$ is
\beq
x_0 = A\cos\omega(t-t_0)\label{4.008}
\enq
where the initial condition has been absorbed wholly in $x_0$. For $x_n$, $n\neq 0$, we need $x_n(t=0)=0$. At the order $n=1$,
\be
\ddot{x}_1+\omega^2 x_1 &=& -x_0^3 = -A^3\cos^3\omega(t-t_0)\nonumber\\
&=&-\f{A^3}{4} \left(3\cos\omega (t-t_0)-\cos 3\omega(t-t_0)\right)\label{4.009}
\en
It is $\cos\omega(t-t_0)$, the resonating term on the right hand side of Eq. (\ref{4.009}) that causes $x_1(t)$ and hence the perturbation theory, to diverge as $t-t_0\rightarrow\infty$. The solution for $x_1(t)$ is
\be
x_1(t) &=& -\f{3A^3}{8\omega}(t-t_0)\sin\omega(t-t_0)\nonumber\\
&& +\f{A^3}{32\omega^2}\left[\cos 3\omega(t-t_0)-\cos\omega(t-t_0)\right]\label{4.010}
\en
giving
\be
x(t) &=& A\cos\omega(t-t_0)-\frac{3\lambda A^3}{8\omega}(t-t_0)\sin\omega(t-t_0)\nonumber\\
&+& \frac{\lambda A^3}{32{\omega}^2}\left[\cos 3\omega(t-t_0)-\cos\omega(t-t_0)\right]+\mc{O}(\lambda^2)\phantom{uuu}\label{4.011}
\en
The $t-t_0$ in the divergence causing term is now split as $t-\tau+\tau-t_0$ where $\tau$ is an arbitrary time and we introduce two renormalization constants as $A=A(\tau)\mc{Z}_1(\tau,t_0)$ and $\theta_0=-\omega t_0=\theta(\tau)+\mc{Z}_2(\tau,t_0)$. The constants $\mc{Z}_{1,2}$ have the expansions
\be
\mc{Z}_1 &=& 1+\lambda a_1+\lambda^2 a_2+\dots \label{4.012}\\
\mc{Z}_2 &=& \lambda b_1+\lambda^2 b_2+\dots \label{4.013}
\en
and we can write Eq. (\ref{4.011}) as
\be
x(t) &=& A(\tau)\left[1+\lambda a_1+\dots\right]\cos\left(\omega t+\theta+\lambda b_1\right)\nonumber\\&&-\f{3\lambda A^3}{8\omega}\left(t-\tau+\tau-\t_0\right)\sin\left(\omega t+\theta\right)+\dots\nonumber\\
&=& A(\tau)\cos\left(\omega t+\theta\right)+\lambda A(\tau) a_1\cos\left(\omega t+\theta\right)\nonumber\\&&- A(\tau)\lambda b_1\sin\left(\omega t+\theta\right)-\f{3\lambda A^3}{8\omega}\left(t-\tau+\tau-\t_0\right)\nonumber\\&& \times\sin\left(\omega t+\theta\right)+\dots \label{4.014}.
\en
Choosing $a_1=0$ and $b_1=-\f{3\lambda A^2}{8\omega}(\tau-t_0)$ we remove the divergence causing terms (the terms containing $\tau-t_0$) and we are left with
\be
x(t)=A(\tau)\cos\left(\omega t+\theta\right)-\f{3\lambda A^3}{8\omega}(t-\tau)\sin\left(\omega t+\theta\right). \phantom{uuu}\label{4.015}
\en
Since the final solution cannot depend on the arbitrary time scale $\tau$, we impose the condition $\left(\f{\partial x}{\partial \tau}\right)_t=0$, for all $t$, which leads to
\be
\f{dA}{d\tau} &=& 0 \label{4.016},\\
\f{d\theta}{d\tau} &=& \f{3\lambda A^2}{8\omega} \label{4.017}.
\en
This is in accordance with what should have happened. The condition, $dA/d\tau=0$, implies a center type periodic orbit and $d\theta/d\tau=g(A)$ implies an amplitude dependent time-period. We want to draw attention to the fact that $dA/d\tau=0$ is a consequence of the absence of $\sin \omega(t-t_0)$ on the right hand side of the Eq.(\ref{4.009}). It will continue to remain zero at $\mc{O}(\lambda^2)$ and above if the right hand side of $\ddot{x}_2+\omega^2x_2$, $\ddot{x}_3+\omega^2x_3$ and so on never has $\sin\omega (t-t_0)$ term. On the other hand, $d\theta/d\tau$ is determined entirely by the coefficient of $\cos\omega (t-t_0)$ on the right hand side of the Eq. (\ref{4.009}). If this term were absent, then $d\theta/d\tau = 0$, which implies an amplitude independent time-period and hence we would have an isochronous oscillation. Thus the signature of isochronous oscillations, $dA/d\tau=0$ and $d\theta/d\tau=0$, is produced by the absence of any driving term involving $\sin\omega (t-t_0)$ and $\cos\omega (t-t_0)$ on the right hand side of the successive differential equations for $x_n(t)$. The vanishing of $dA/d\tau$ and $d\theta/d\tau$ gives an isochronous center or else it gives at every order a relation among the constants in the problem that has to be satisfied for an isochronous orbit.
\\
\section{Examples}
\subsection{A simple case}
As a first check that this statement is valid, we considered the most well known nontrivial isochronous system --- the oscillations in the potential $V(x)=\f{1}{2}\left(x^2+\f{1}{x^2}\right)$. The equation of motion is given by
\beq
\ddot{x}=-x+\f{1}{x^3}\label{4.018}
\enq
The linear center is clearly at $x=\pm 1$ and it suffices to study the orbit around only one of them (there is an infinite barrier at $x=0$ and hence the two centers are completely decoupled). We shift the origin to the center according to our strategy and write eq. (\ref{4.018}), in terms of the new variable $y=x-1$,
\be
\ddot{y} &=& -(1+y)+(1+y)^{-3}\nonumber\\
&=& -4y+6y^2-10y^3+15y^4-21y^5+\dots \label{4.019}
\en
For book keeping purposes we introduce a parameter $\lambda$ and rewrite eq. (\ref{4.019}) as
\beq
\ddot{y}+4y=\lambda 6 y^2-\lambda^2 10 y^3+\lambda^3 15 y^4-\lambda^4 21 y^5+\dots \label{4.020}
\enq
Now we can expand $y$ as
\beq
y=y_0+\lambda y_1+\lambda^2 y_2+\lambda^3 y_3+\dots \label{4.021}
\enq
and proceed as earlier. At different orders of $\lambda$ we find the following differential equations:
\be
\lambda^0:\quad\quad \ddot{y}_0+4 y_0 &=& 0 \label{4.022},\\
\lambda^1:\quad\quad \ddot{y}_1+4 y_1 &=& 6 y_0^2 \label{4.023},\\
\lambda^2:\quad\quad \ddot{y}_2+4 y_2 &=& 12 y_0 y_1-10 y_0^3 \label{4.024},\\
\lambda^3:\quad\quad \ddot{y}_3+4 y_3 &=& 12 y_0 y_2+6 y_1^2-30 y_0^2y_1 +15 y_0^4\label{4.025}\phantom{uuu}
\en

and so on. We work with the initial conditions $y=A_0$, $\dot{y}=0$ at $t=0$. We let $x_0$ pick up the initial condition ($y_0=A_0$, $\dot{y_0}=0$), so that for all subsequent orders --- $x_i(t=0)=\dot{x}_i(t=0)=0$. Explicit calculations upto 6th order in $\lambda$ confirms the absence of resonating terms ($\cos\Omega t$ or $\sin\Omega t$) in the inhomogenous part of the equations at each order, establishing that upto $\mc{O}(\lambda^6)$, both $dA/d\tau=0$ and $d\theta/d\tau=0$ and that our condition (\ref{4.003}) for isochrony is satisfied by this well known center type isochronous oscillator.
\\
\subsection{Cherkas System}
We now turn to a well studied and rather interesting example --- a dynamical system introduced by Cherkas. This system was first studied by Cherkas \cite{cher} but the center conditions found therein were later found to be incomplete. The complete set of necessary and sufficient conditions were found in \cite{clp} and \cite{lcdpy}. The conditions for which the origin is an isochronous center was found by \cite{hlp}. We deal with this system in some detail, so that our methodology becomes transparent. The equation has the form
\be
\dot{x} &=& y(1+x), \label{i4.032a}\\
\dot{y} &=& -x-a_1x^2-a_2x^3-a_3x^4-a_4(x+a_5x^2)y-a_6y^2.\phantom{uuu} \label{i4.032b}
\en
We can write this as a second order differential equation,
\be
\ddot{x} &=& (1+x)\dot{y}+y \dot{x}, \nonumber\\
&=& -(1+x)\left[x+a_1 x^2+a_2 x^3+a_3x^4+a_4(x+a_5x^2)y+a_6y^2\right]
\nonumber\\ &&+\f{\dot{x}^2}{1+x}(1-a_6), \label{i4.033}
\en
which we cast as
\be
\ddot{x} + x &=& -(1+a_1)x^2-(a_2+a_1)x^3-(a_2+a_3)x^4-a_3x^5\nonumber\\ &&-a_4 \dot{x}(x+a_5x^2)+(1-a_6)\f{\dot{x}^2}{1+x}, \label{4.034a}\\
&=& -\beta x^2-(a_1+a_2)x^3-(a_2+a_3)x^4-a_3x^5\nonumber\\ &&-a_4 \dot{x}(x+a_5x^2) +\alpha \dot{x}^2\left[1-x+x^2-x^3+\dots\right].\phantom{uuu} \label{4.034b}
\en
where $\beta=1+a_1$ and $\alpha=1-a_6$.
The book keeping parameter $\lambda$ that we introduce is distributed as follows
\be
\ddot{x} + x &=& -\lambda \beta x^2-\lambda^2\left(a_1+a_2\right)x^3 -\lambda^3\left(a_2+a_3\right)x^4 \nonumber\\ 
&& -\lambda^4 a_3 x^5  + a_4 \lambda x \dot{x} -a_4a_5\lambda^2 \dot{x}x^2+\alpha \lambda \dot{x}^2\nonumber\\ && \times(1-\lambda x+\lambda^2x^2-\lambda^3x^3+\dots)\label{4.034}.
\en
We now expand $x$ as,
\beq
x=x_0+\lambda x_1+\lambda^2 x_2+\cdots \label{4.035}.
\enq
Using the above expansion in eq. (\ref{4.034}), at different orders, we get

\be
\ddot{x}_0+x_0 &=& 0 \label{4.036},\\
\ddot{x}_1+x_1 &=& -\beta x_0^2+\alpha \dot{x}_0^2 -a_4x_0\dot{x}_0\label{4.037},\\
\ddot{x}_2 + x_2 &=& -2\beta x_0x_1-(a_1+a_2)x_0^3+2 \alpha \dot{x}_1 \dot{x}_0 - \alpha \dot{x}_0^2 x_0 \nonumber \\ &&-a_4(x_1\dot{x}_0+x_0\dot{x}_1)-a_4a_5\dot{x}_0 x_0^2.\label{4.038}
\en
We use the initial conditions $x(t=0)=A$ and $\dot{x}(t=0)=0$ and impose the condition on $x_0$, so that all subsequent $x_n$ have the initial conditions $x_n(0)=\dot{x}_n(0)=0$. Upto $\mc{O}(\lambda)$, the solutions are,
\be
x_0 &=& A_0\cos t \label{4.039},\\
x_1 &=& A_0^2\Big[\frac{\alpha-\beta}{2}+\frac{\alpha+\beta}{6}\cos 2t - \frac{a_4}{6}\sin 2t \nonumber \\ && \phantom{uu}-\frac{2\alpha-\beta}{3}\cos t+\frac{a_4}{3}\sin t\Big] \label{4.040}.
\en
Inserting these solutions in the right hand side of eq. (\ref{4.038}), we find that the condition for the existence of a center (vanishing of the coefficient of the $\sin t$ term) is
\be
&& a_4(\beta-\alpha-a_5) = 0 \label{i4.040a}\\
\textrm{i.e.}\quad && a_4 = 0 \quad \textrm{or} \quad \beta-\alpha =a_5 \label{i4.040b},
\en
and the condition for isochronicity is
\beq
\f{\alpha^2}{3}-\f{\alpha}{4}+\f{a_4^2}{12}=\f{3}{4}(a_1+a_2)+\f{5}{6}\alpha\beta-\f{5}{6}\beta^2\label{4.041}.
\enq
These conditions hold only for $a_3=0$, in the system equation (\ref{i4.033}). This is because the order upto which we have calculated so far does not include the $a_3$ term at all. However, notice that the conditions given above are only necessary conditions. They need to be satisfied for the existence of isochronous center for $a_3=0$, but obviously do not guarantee the existence of a center. If we have a center with even $a_4=0$ (the case considered by Ghose Choudhury and Guha), then for $a_3=0$, the isochronicity condition turns out to be
\beq
\f{5}{6}\alpha(1+a_1)-\f{5}{6}(1+a_1)^2+\f{3}{4}(a_1+a_2) = \f{\alpha^2}{3}-\f{\alpha}{4} \label{i4.042},
\enq
where $1+a_1=\beta$. It is to be noted that all the isochronous cases found with $a_3=0$ satisfy the condition given in eq.(\ref{i4.042}). In the interesting special case of $\alpha=\beta$, this condition reduces to
\beq
\left(\alpha-\f{3}{2}\right)^2=\f{9}{4}a_2. \label{i4.043}
\enq
In the two isochronous cases \cite{guch}, we have encountered so far, with $\alpha=\beta$, the results are in accordance with eq.(\ref{i4.043}). The usefulness of such constraints lie in the fact that in one's search for isochronous systems, one knows in which subspace the search has to be limited. For $a_3\neq 0$, the calculation needs to be carried through two more orders and at the end of a long but straightforward calculation one obtains
\be
&& \f{5}{4}\alpha^4 + \f{65}{48} \beta^4 + \f{355}{48}\alpha^2 \beta^2 -\f{65}{12}\alpha\beta^3-\f{55}{12}\alpha^3\beta -\f{31}{24}\alpha^3 \nonumber\\ && - \f{5}{12}\alpha\beta^2 
+ \f{83}{48} \alpha^2\beta -\f{39}{16}\beta^2(a_1+a_2)+\f{69}{12}\alpha\beta(a_1+a_2) \nonumber\\ &&- \f{27}{8} \alpha^2(a_1+a_2) 
-\f{1}{3} \alpha\beta +\f{83}{192} \alpha^2 + \f{7}{4} \beta (a_2+a_3) \nonumber\\ && -\f{3}{2}\alpha(a_2+a_3) -\f{1}{16}\alpha(a_1+a_2) +\f{3}{64}(a_1+a_2)^2 \nonumber\\ &&  -\f{5}{8}a_3 - \f{\alpha}{8} = 0\label{i4.044}
\en
Once again all the $a_3\neq 0$ cases noted by Ghose Choudhury and Guha satisfy the above constraint.\\
When $a_3 = 0$ and $\alpha = \beta$ the above constraint takes on a much simpler form given by,
\be
&& \f{\alpha^3}{48} - \f{1}{16}\alpha^2 \left(a_1 + a_2\right) + \f{19}{192} \alpha^2 + \f{1}{4} \alpha a_2 - \f{1}{16} \alpha \left(a_1 + a_2\right) \nonumber\\
&& \phantom{uuuuuuuuuuuuuuuu} + \f{3}{64} \left(a _1 + a_2 \right)^2 - \f{\alpha}{8} = 0. \label{i4.044}
\en
For Cherkas system when $a_3 = 0$ and $\alpha = \beta$ all the isochronous systems must satisfy the above condition. Keeping in mind the fact that $\alpha = \beta = 1 + a_1$, the above constraint gives a relation between $a_1$ and $a_2$ much like the constraint obtained at previous order, given by the Eq. (\ref{i4.043}). So we have two equations for two parameters which must be satisfied for isochrony and thus the two constraints ( Eqs. (\ref{i4.043}) and (\ref{i4.044})) constitute both {\it necessary} and {\it sufficient} conditions for an isochronous Cherkas system, with $a_3 = 0$ and $\alpha = \beta$. Now eliminating $a_2$ between Eqs. (\ref{i4.043}) and (\ref{i4.044}) we are left with a cubic equation in $\alpha$, given by
\beq
4 \alpha^3 -24 \alpha^2 + 45 \alpha - 27 = 0 \label{i4.045}.
\enq
There are only two roots of the above cubic equation  i.e. $\alpha = 3$ and $\alpha = 3/2$ (which is a repeated root). These two cases corresponding to the parameter values of: $\alpha = 3$, $a_2 =1$ and  $\alpha = 3/2$, $a_2 =0$ are the only possible values for which Cherkas system (when $a_3 = 0$ and $\alpha=\beta$) is isochronous.
\\
To further establish the efficacy of our method we consider a set of cubic systems as defined by Bardet et al given by,
\be
\dot{x} &=& -y(1+a_1 x) + a_2 x^2 + a_3 x^3, \label{i4.045a}\\
\dot{y} &=& x + b_3 x^2 + b_5 x^3 - y(b_2 x - b_4 x^2)- b_1 y^2. \label{i4.045b}
\en
The necessary condition for the isochronous centers as found by us is
\be
&& (2a_2-b_2)\left[\f{a_1+b_1}{3}-\f{b_3+a_1}{6}+\f{3}{4}b_3-\f{b_1}{4}+\f{a_1}{2}\right]\nonumber\\
&& +a_2(2b_1+a_1)-b_4-3a_3 = 0 \label{i4.046}
\en
and
\be
&& \f{(a_1+b_1)}{3}(b_3+b_1+2a_1)(b_3+b_1+2a_1)-\f{b_3+a_1}{6}\nonumber\\ && \times(7b_1-5b_3+2a_1) -\f{3}{4}(b_5+a_1b_3-a_2b_2)\nonumber\\ &&-\f{a_1(a_1+b_1)}{4}+\f{1}{12}(2a_2-b_2)^2 = 0 \label{i4.047}
\en
All the system mentioned in Theorem 4.1 of Bardet et al satisfy the criterion given above.
\subsection{Ricatti Equation}
We finally turn to a variant of the Ricatti system of Type-II, given by
\beq
\ddot{x}+kx\dot{x}+\f{k^2}{9}x^3+\alpha x_1 = 0 \label{4.042}.
\enq
Oscillators having equation of motion of the form $\ddot{x}+f(x)\dot{x}+g(x)=0$ are called Li$\acute{\textrm{e}}$nard type oscillators. For our example $f(x)=kx$ and $g(x)=({k^2}/{9})x^3+\alpha 
x_1$. This oscillator has been widely treated in the literature and has been referred to as with various names. The oscillator may be interpreted as a cubic anharmonic oscillator with a damping type nonlinear force $kx\dot{x}$. The equation has otherwise been called the generalized Emden-type equation \cite{leach} and has been widely studied in the literature over the last two decades. The equation arises in a variety of physical problems --- like the modeling fusion of pellets \cite{erw}, one dimensional analogue of Yang-Mill's boson gauge theory \cite{chicom}, equilibrium configurations of a spherical cloud subject to the laws of thermodynamics etc. \cite{scnpra}.  Recently Chandrasekhar et al \cite{vkcsl} have shown that for $\alpha>0$, this oscillator admits non-isolated periodic orbits with the unusual property that frequency of oscillation remains independent of amplitude and same as that of the linear oscillator. For our purposes we will show that the isochronous periodic solution that the eq. (\ref{4.042}) admits, also follows the criterion we have put forward in this paper, i.e. $dA/d\tau=d\theta/d\tau=0$. Assuming that $k$ is small we expand $x$ in powers of k as follows:
\beq
x=x_0 + k x_1 + k^2 x_2 + k^3 x_3\dots \label{4.043}
\enq
We work with the initial condition as $x(t_0)=A$ and $\dot{x}(t_0)=0$ and absorb the initial condition wholly at zeroth order as we had done earlier. Proceeding henceforth according to our prescription we calculate the flow equations perturbatively. We find, as per our expectations the flow equations turn out to be $dA/d\tau=d\theta/d\tau=0$, right upto order $\mc{O}(k^3)$ and we have an indication of an isochronous type oscillator. Numerical simulations of the oscillator represented by eq. (\ref{4.042}), reveal that for values of $\alpha(=\Omega^2)>1$, one obtains isochronous non-isolated orbits and so is the conclusion of the work by Chandrasekhar et al \cite{vkcsl}. Using our RG approach we have shown here that for a oscillator to execute isochronous oscillations it must obey the criterion we have proposed, i.e. both $dA/d\tau$ and $d\theta/d\tau$ must be zero.
\\
\section{Conclusion}
The above example is a rather interesting case where even after addition of nonlinearity the oscillator frequency remains independent of amplitude and same as that of the linear harmonic oscillator. This opens up the possibility of identifying more such nonlinear oscillators which admit oscillations with amplitude-independent frequency. One can look to identify suitable nonlinearities which, added to the linear harmonic oscillator case, do not generate resonating terms at different orders of perturbation. Thus our method can in principle provide the necessary conditions for finding oscillators having this property. Moreover we can ask the question what is the most general case of a nonlinear oscillator of the form $\ddot{x} = -x + f(x,\dot{x})$ that can admit solutions exhibiting isochronous center-type oscillations. We believe our method has important implications in developing nonlinear systems exhibiting isochronous oscillations. The obvious limitation of our methodology, is being perturbative, it provides only the necessary and not necessarily sufficient conditions for isochronicity. However due to its simplicity it allows us to get an immediate result for whether a particular set of nonlinearities will lead to isochronicity or not and thus also indicates the tinkering necessary to get an isochronous system. Further this method can serve as a robust way to perturbatively construct the so called \textit{period functions} which must vanish for isochrony. Much of the work in the field of isochronous oscillations involves finding a Hamiltonian description of the system and going on to find conditions for integrability or super-integrability. However the technical difficulties involved have prevented a complete solution (e.g. for the Cherkas system) at one shot. Besides many cases can't be immediately reducible to Hamiltonian systems and in most cases it proves rather difficult to find first integrals when nonlinearities are involved. Our method however is not limited by such considerations and can be used effectively as a first probe for isochronous dynamical systems.
\\



\end{document}